# Game Theory and Coverage Optimization Based Multihop Routing Protocol for Network Lifetime in Wireless Sensor Networks

Yin-Di Yao, Xiong Li, *Student Member, IEEE*, Yan-Peng Cui, *Student Member, IEEE*, Lang Deng, and Chen Wang

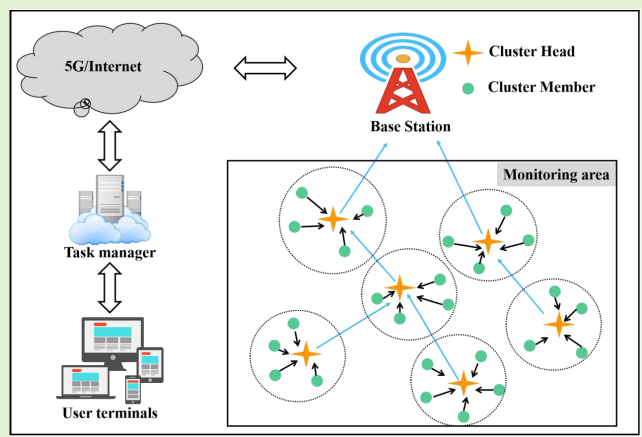

*Abstract*—Wireless sensor networks (WSNs) are self-organizing monitoring networks with a large number of randomly deployed micro sensor nodes to collect various physical information so as to realize tasks such as intelligent perception, efficient control and decision-making. However, WSNs nodes are powered by batteries, so they will run out of energy after a certain time. This energy limitation will greatly constrain the network performance like network lifetime and energy efficiency. In this study, for the purpose of prolonging network lifetime, we proposed a multi-hop routing protocol based on game theory and coverage optimization (MRP-GTCO). Briefly, in the stage of setup, two innovational strategies including clustering game with penalty function and cluster head coverage set were designed to realize the uniformity of cluster head distribution and improve the rationality of cluster head election. In the data transmission stage, we firstly derived the applicable conditions theorem of inter-cluster multi-hop routing. Based on this, a novel multi-hop path selection algorithm related to residual energy and node degree was proposed to provide an energy-efficient data transmission path. The simulation results showed that MRP-GTCO protocol can effectively reduce the network energy consumption and extend the network lifetime 159.22%, 50.76% and 16.46% compared with LGCA, RLEACH and ECAGT protocols.

*Index Terms*—Coverage optimization, game theory, network lifetime, routing protocol, wireless sensor networks.

## I. INTRODUCTION

WIRELESS sensor networks have brought a new revolution to information sensing [1] because of its advantages such as convenient deployment, strong anti-destructive ability and low cost, and have been widely applied in intelligent agriculture [2], environmental monitoring [3] and military monitoring [4], etc. It is a self-organizing network composed of low-power micro sensor nodes, which can perform wireless communication, data sensing, processing and storage for IoT applications. However, sensor nodes are usually powered by batteries with limited capacity [5] and deployed in harsh natural environments [6], which makes it difficult to charge, let alone replace batteries. Once the node dies because of energy exhaustion, it will bring irreparable disasters to the whole network such as incomplete monitoring data and great changes of network topology will lead to decision delay or error. Generally, network lifetime is reflected by the round of dead nodes and the most popular definitions are the rounds of the first dead node (FDN), half dead nodes (HDN) and last dead node (LDN) [7], [8]. Network energy consumption determines the death rate of nodes, and then affects the network lifetime. Therefore, it is of great significance to reduce the energy consumption of nodes and prolong the network lifetime.

Long-distance data transmission, data reception and fusion are the main energy consumption sources of nodes. Note that the routing protocol in the network layer is mainly concerned with data reception, occurrence and path selection. Therefore,

Manuscript received May 1, 2022; accepted May 24, 2022. Date of publication June 3, 2022; date of current version July 1, 2022. This work was supported in part by the National Natural Science Foundation of China under Grant U1965102, in part by the Science and Technology Innovation Team for Talent Promotion Plan of Shaanxi Province under Grant 2019TD-028, and in part by the Science and Technology Department of Shaanxi Province Project under Grant 2021NY-180. The associate editor coordinating the review of this article and approving it for publication was Prof. Jaime Lloret. *(Corresponding author: Xiong Li.)*

Yin-Di Yao, Xiong Li, Lang Deng, and Chen Wang are with the School of Communication and Information, Xi'an University of Posts and Telecommunications, Xi'an 710121, China, and also with the Shaanxi Key Laboratory of Information Communication Network and Security, Xi'an 710121, China (e-mail: yaoyindi@xupt.edu.cn; leexiong123@163.com; 951854473@qq.com).

Yan-Peng Cui is with the School of Information and Communication Engineering, Beijing University of Posts and Telecommunications, Beijing 100876, China (e-mail: cuiyanpeng94@bupt.edu.cn).

designing an energy-efficient routing protocol is a key issue to prolong the network lifetime. Clustering routing protocol is envisioned as a promising solution [6], [9], becoming one of the research hotspots of WSNs routing protocol because of its strong scalability, low load and energy consumption, and support for data fusion. In the cluster routing protocol, all nodes are divided by clusters, and a certain node in the cluster is elected as the cluster head. On the one hand, it is responsible for coordinating the monitoring and data transmission tasks of the cluster members in the cluster, and on the other hand, it forwards data to the base station in a multi-hop manner. However, due to undertaking the task of data fusion and forwarding, cluster heads often run out of energy earlier than other nodes. In addition, unreasonable relay forwarding paths will lead to excessive load and unbalanced energy consumption of relay cluster heads, which will further accelerate the death of cluster heads and dramatically shorten the network lifetime. Therefore, how to optimize (relay) cluster head selection to prolong network lifetime is one of the crucial challenges facing clustering routing protocols.

### A. Related Work and Motivation

In recent years, many scholars have applied game theory to the construction of clustering routing protocol in wireless sensor networks [10], [11], successfully reducing network energy consumption and prolonging network lifetime.

Koltsidas G *et.al* presented a game theroy based protocol called Clustered Routing for Selfish Sensors (CROSS) [12], where game theory was used to model the cluster head selection for wireless sensor networks. Each node is modeled as a player and the payoffs in the cluster head selection game depend on whether the node declares itself a cluster head. The author also proved the existence of a Nash equilibrium, where no node has the incentive to change the balance if every node declares itself a cluster head with probability $p = 1 - w^{\frac{1}{N-1}}$. Here, $N$ is the number of nodes and $w$ is a predefined parameter related to payoff and cost of nodes. Sadly, CROSS regards the clustering game global way, namely every sensor can hear the transmissions from all nodes and compete cluster head select with all nodes which is not realistic because every sensor node has a certain transmission distance due to limited power. Worse, in the CROSS protocol, it is very likely that there is no cluster head in some rounds, which makes it impossible to efficiently forward data to the base station. In addition, in the process of probability calculation, the parameter $w$ of CROSS is obtained through simulation experiments, without considering the actual energy consumption cost of nodes, so it is difficult to achieve its optimal performance.

To overcome these problems, scholars have made a lot of improvements to the CROSS protocol. Chen *et.al* [13] embodies the node parameter $w$ and re-derives the Nash equilibrium solution $p = 1 - (\frac{c_D - c_{RD}}{v - c_{RD}})^{\frac{1}{N-1}}$. Where $c_D$ and $c_{RD}$ represent the cost of the node when it declares itself as cluster head and the refusing situation, respectively. However, the second selection of cluster heads based on clustering game is not considered, which leads to redundancy and low utilization rate of cluster heads. Xie *et.al* [14] proposed a Localized game theoretical clustering algorithm (LGCA) where each node only compet with its neighbour node within a communication radius $R_C$ and a final cluster head contention mechanism was developed to to ensure that only one cluster head in a certain region. Although LGCA can avoid the huge cost brought by the global game, there are still many aspects that have not been considered, such as residual energy, node degree, etc. Hence, it is difficult to effectively improve the rationality of cluster head selection. Hybrid game theory–based and distributed clustering (HGTD) [15] was a clustering protocol proposed on the basis of LGCA. It defines the payoff for each node when choosing different strategies, where both node degree and distance to base station are considered in the process of the localized clustering game. Meanwhile, for ensuring the uniformity of cluster head distribution, the tentative cluster head with more residual energy and fewer neighbor cluster heads has a bigger probability to be the final cluster head. However, the position of cluster head is the key factor that affects the uniformity of cluster head distribution, and the author ignores this point, which leads to no essential improvement of uniformity. In [16], Energy-efficient clustering algorithm based on game theory (ECAGT) was proposed by adding the residual energy factor into the equilibrium probability $p = 1 - w^{\frac{1}{N_i - 1}} \times (\frac{E_i}{E_{ave}})^\alpha$, where $E_i$ is the residual energy of the node $i$. $E_{ave}$ is the average energy of all the nodes within sensor node $i$'s communication radius. For final cluster head selection, only the remaining energy of candidate cluster heads is considered, and the actual network conditions such as node degree and distribution of candidate cluster heads are ignored.

Although the above routing protocol has successfully applied game theory to the clustering routing protocol and considerable performance improvement has been achieved, there are still many problems to be solved. For example, the above routing protocols focus on solving the problem of cluster head selection and uniform distribution, but ignore the influence of the actual deployment of network nodes on the design of the clustering game such as location and load of candidate cluster heads, etc. Specifically, for some sub-regions with dense nodes, at least one cluster head should be guaranteed to undertake forwarding tasks to reduce network energy consumption. For this type of regional nodes, if they do not undertake cluster heads in the clustering game, they should be punished for becoming cluster heads. In addition, the nodes with more residual energy should be given greater probability to act as cluster heads in the game, because they are not easy to die when acting as cluster heads, compared with low-energy nodes. Moreover, when selecting cluster heads, the position distribution among cluster heads is ignored, which is the primary reason to ensure the uniformity of cluster heads. Lastly, to our best knowledge, the existing clustering routing protocols based on game theory are all single-hop routing protocols, and their performance will be greatly reduced when they perform tasks in large monitoring areas. Table I summarizes differences and similarities of the related work.

TABLE I
SUMMARY OF THE RELATED WORK

| Protocol | Clustering game and consideration factor | Factor of final CH selection | Routing between cluster heads |
|---|---|---|---|
| CROSS[12] | Global game; $p = 1 - w^{\frac{1}{N-1}}$ | —— | One-hop routing |
| Literature[13] | Global game; $p = 1 - (\frac{c_D - c_{RD}}{v - c_{RD}})^{\frac{1}{N-1}}$ | —— | One-hop routing |
| LGCA[14] | Local game; $p = 1 - w^{\frac{1}{N_b - 1}}$ | CSMA/CA mechanism | One-hop routing |
| HGTD[15] | Local game; $p = 1 - w^{\frac{1}{N_b - 1}}$ | Residual energy and the number of neighbor CHs | One-hop routing |
| ECAGT[16] | Local game; $p = 1 - w^{\frac{1}{N_i - 1}} \times (\frac{E_i}{E_{ave}})^\alpha$ | Residual energy | One-hop routing |
| Our proposed | Local game; $p_i = 1 - (\frac{\Phi E_{CH} - E_{CM}}{\Phi E_{CH}})^{\frac{1}{N_{bS_i} - 1}}$ ( **Novel penalty coefficient** $\Phi$) | Residual energy and **cluster head position** | **Multi-hop routing** |

## B. Main Contributions

For the purpose of addressing the above problems, in this work, we propose a multi-hop routing protocol based on game theory and coverage optimization (MRP-GTCO) to realize the near-optimal performance on cluster head and multi-hop path selection. The major contributions of this paper are concluded as follows.

1) To avoid that the data of cluster members cannot be forwarded because the selfishness of nodes does not become the cluster head in current network, a punishment mechanism related to node residual energy and node degree is designed. The probability of nodes becoming cluster heads in Nash equilibrium of clustering game with penalty mechanism is analyzed.
2) By constructing the relationship between the coverage optimization and the uniform distribution of cluster heads, an innovative cluster head selection algorithm based on the cluster head coverage rate and residual energy is proposed to realize the uniform distribution of cluster heads so as to reduce the communication energy consumption among nodes (cluster head and cluster members or cluster head and destination node).
3) Aiming at transmitting intra-cluster data to base stations with the least energy consumption by cluster heads in different scenarios, the optimal multi-hop relay selection theorem between clusters is constructed. Based on the above theorem, a novel relay node selection algorithm related to the residual energy of nodes and the number of neighboring nodes is proposed greatly reducing the network energy consumption caused by long-distance data forwarding.

The rest of the paper is structured as follows: the assumptions of system model, energy consumption model and node consumption analysis are introduced in Section II. Section III describes and analyzes the proposed routing protocol. In Section IV, the simulation results and the causes of performance differences are analyzed in detail. At last, we summarized the research of this paper and proposed the future work. In addition, the key parameters used in this paper are listed in Tables II, respectively.

## II. PRELIMINARIES

### A. Network Model

Assume that there are $N$ sensor nodes are randomly deployed in $M \times Mm^2$ region. We make the following assumptions to facilitate our proposed routing protocol analysis.

1) All sensor nodes have limited and equal initial energy, and the processing and communication capacity are the same. But the energy of the base station is unlimited.
2) The sensor nodes are randomly deployed in the monitoring area and the position of sensor nodes is fixed after deployment.
3) The base station is located in the centre of the monitoring area.
4) Sensor nodes can obtain their location through GPS or other localization methods such as DV-Hop, TDOA, etc.

TABLE II
KEY PARAMETERS USED IN THIS PAPER

| Parameters (unit) | Notation |
|---|---|
| $E_{mp}$ $(pJ/bit/m^4)$ | Multi fading energy consumption coefficient |
| $E_{fs}$ $(pJ/bit/m^2)$ | Free space energy consumption coefficient |
| $E_{elec}$ $(nJ/bit)$ | Transceiver circuit consumption coefficient |
| $E_{DA}$ $(nJ/bit)$ | Data fusion energy consumption coefficient |
| $d_o$ $(m)$ | Distance thresholds of energy consumption models |
| $l$ $(bits)$ | The size of Data packet |
| $K_{opt}$ (nodes) | Optimal number of cluster heads |
| $M$ $(m)$ | Monitoring area side length |
| $U(S_i)$ $(J)$ | Energy utility function of $S_i$ |
| $\Phi(S_i)$ | Penalty coefficient related to energy and node degree |
| $\Gamma_{total}$ | Coverage rate of cluster heads |
| $E_R(J)$ | Energy consumption of relay strategy |
| $E_{NR}(J)$ | Energy consumption with non-relay strategy |
| $E_{CR}(J)$ | Energy consumption of the cluster head as a relay node |

### B. Energy Consumption Model

The energy of the senosor node is mainly consumed in the communication. We use the popular energy dissipation model [17] to calcuate energy consumption of senseor nodes. The energy consumed in the transmission of $l$-bit data at the distance $d$, which is represented by $E_{Tx}$, shown in (1). If $d \leq d_o$, the free space model is used and $\varepsilon_{fs}$ is the energy factor per bit in the free space model; otherwise, the multipath fading model is used and $\varepsilon_{mp}$ is the energy factor per bit in multipath fading model. The threshold $d_o$ can be computed by (2). $\varepsilon_{elec}$ is the energy factor to send or recieve per bit.

$$E_{Tx}(l,d) = \begin{cases} l \times \varepsilon_{elec} + l \times \varepsilon_{fs} \times d^2, & if \ d \leq d_o \\ l \times \varepsilon_{elec} + l \times \varepsilon_{mp} \times d^4, & if \ d > d_o \end{cases} \quad (1)$$

$$d_o = \sqrt{\varepsilon_{fs}/\varepsilon_{mp}} \quad (2)$$

$E_{Rx}$ represents the energy comsumed by recceiving $l$-bit data, and its expression is

$$E_{Rx}(l) = l \times \varepsilon_{elec}. \quad (3)$$

### C. Node Energy Consumption and Optimal Cluster Head Number

To facilitate the subsequent theoretical derivation, we first analyze the energy consumption sources of cluster head and cluster member, and then deduce their mathematical formulas. Furthermore, the expressions of energy consumption of one cluster and the total network are derived from cluster head and cluster member energy consumption.

For the multihop clustering routing protocol, the cluster head energy consumption consists of the following three parts:

1) **(I)** to receive the sensing data from $N/K - 1$ cluster members and from other $N_r$ cluster heads' $N_C = N_r \times N_c$ sensing data. Where, $K$ is the number of cluster heads and $N_c$ indicates the average number of forwarded packets from other $N_r$ cluster heads.
2) **(II)** to fuse $N/K$ data from cluster members. Note that because the data from other cluster heads are generally not similar to the data of their own clusters, the cluster head only fuse the data in their own clusters.
3) **(III)** to send this data to the destination node, which may be the relay node or base station.

Thanks to the cluster heads being uniformly distributed in the monitoring area under the help of our cluster head selection mechanism that will be introduced in the next section in detail, the communication distance between cluster head and its destination node can conform to the free space model. Therefore, the energy consumption of a cluster head can be expressed by (4). Here, $d_{toDest}$ represents the distance between the cluster head and the destination node. Here, $\varepsilon_{DA}$ is the energy factor of fusing per bit data.

$$E_{CH} = l\varepsilon_{elec}(\frac{N}{K} - 1) + l\varepsilon_{DA}\frac{N}{K} + \cdots \\ + (1 + N_C)l \times (l\varepsilon_{elec} + l\varepsilon_{fs}d_{toDest}^2) \quad (4)$$

For cluster members, their energy consumption includes sending the sensing data to the cluster head as shown in (5). Here, $d_{CMtoCH}$ represents the distance between the cluster head and the cluster member. Generally speaking, cluster members are close to their cluster head, so the energy dissipation follows the free space model.

$$E_{CM} = lE_{elec} + l\varepsilon_{fs}d_{CMtoCH}^2 \quad (5)$$

The area occupied by each cluster is approximately $M^2/K$. Cluster members are uniformly distributed and the distribution density function is $\rho(x,y) = K/M^2$. If we assume this cluster area is a circle with radius $R = M/\sqrt{\pi K}$. The expected squared distance from cluster members to the cluster head (assumed to be at the center of mass of the cluster) is given by (6).

$$E[d_{CMtoCH}^2] = \iint (x^2 + y^2)\rho \, dydx \\ = \frac{K}{M^2}\int_0^{2\pi}\int_0^R r^3 dr d\theta = \frac{M^2}{2\pi K} \quad (6)$$

Each cluster area includes one cluster head node and $N/K - 1$ cluster members, so the energy consumption of each cluster can be calculated by

$$E_{Cluster} = E_{CH} + (\frac{N}{K} - 1)E_{CM} \approx E_{CH} + \frac{N}{K}E_{CM}, \quad (7)$$

and it is not difficult to get that the energy consumption of a network with $K$ clusters is

$$E_{Network} = KE_{Cluster} = Nl\varepsilon_{DA} + \frac{\varepsilon_{fs}lNM^2}{2\pi K} \\ + \ldots (2N + KN_C)l\varepsilon_{elec} + Kl(N_C + 1)\varepsilon_{fs}d_{toDest}^2. \quad (8)$$

Based on node energy consumption analysis, we can conclude that the network energy consumption is affected by the number of cluster heads ($K$), because other parameters are constant. Therefore, we can minimize network energy consumption related to the number of cluster heads by setting the derivative of $E_{Network}$ with respect to $K$ to zero. The optimal number of cluster heads can be computed by (9).

$$K_{opt} = M\sqrt{\frac{\varepsilon_{fs}N}{2\pi[\varepsilon_{elec}N_C + \varepsilon_{fs}(N_C + 1)d_{toDest}^2]}} \quad (9)$$

*Remark 1:* Note that with the operation of the network, the number of surviving nodes ($N_{alive}$) in the network will be less than the total number of nodes $N$. However, $K_{opt}$ should also adaptively and dynamically change with the number of surviving nodes ($N_{alive}$). So the optimal number of cluster heads can be also computed by (10).

$$K_{opt} = M\sqrt{\frac{\varepsilon_{fs}N_{alive}}{2\pi[\varepsilon_{elec}N_C + \varepsilon_{fs}(N_C + 1)d_{toDest}^2]}} \quad (10)$$

## III. THE PROPOSED PROTOCOL

In this section, we will present the MRP-GTCO protocol, which includes the following phases: cluster head selection, cluster formation and data transmission. A complete round consists of the above three phases and the operation of our protocol is partitioned into rounds [16]. Fig. 1 shows the overall operation of MRP-GTCO protocol.

Now, we introduce the three phases of the MRP-GTCO protocol respectively, including clustering game based candidate cluster head selection algorithm, final cluster head selection algorithm for coverage optimization and multi-hop routing mechanism in data transmission phase. Before introducing the phase of cluster head selection, we first derive the Nash equilibrium expressions in clustering game that is the probability the node acts as a cluster head.

### A. Clustering Game

Clustering mechanism is an effective method to improve the energy efficiency of the overall wireless sensor network. Importance factors related to energy and physical characteristics of the nodes are used to evaluate the rationality of cluster selection in various clustering mechanisms [18]. However, each node in the network tends to avoid being a cluster head to preserve energy. But additionally, it needs to provide the required services. So the nodes in the network behave selfishly, meaning that their primary goal is to maximize their own benefits (minimize energy consumption) while minimizing their own contribution for the benefit of other nodes. Therefore, through the mutual game and intelligent interaction between nodes, a game theory based clustering mechanism can be formed, thus improving network energy efficiency.

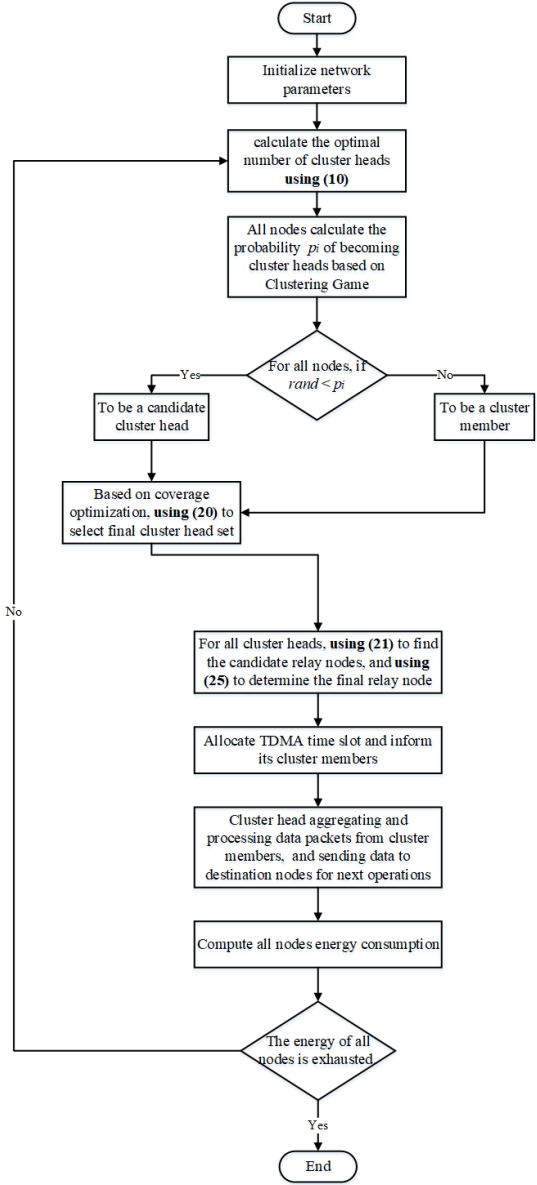

Fig. 1. Workflow of MRP-GTCO protocol.

We first define a game theory based clustering game ($CG$) with a selfish punishment mechanism to optimize cluster head selection. The clustering game is presented as $CG = \{N, S, U\}$. The players are $N$ sensor nodes and each sensor node has the same action/strategy space $A$. When the set of strategies of a node is to decide to be a cluster head (CH) or a cluster member (CM), it is denoted as $A = \{CH, CM\}$, and $U_i$ is utility functions of the nodes after selecting its action from $A$.

Obviously, there are two strategies for every sensor nodes:
1) If all nodes don't want to be cluster heads, this will cause all nodes to transmit data directly to the base station, which will lead to greater network overhead. At this time, the payoff of all nodes are 0.
2) If the existing node is willing to become the cluster head, the income of the node that becomes the cluster head at this time is defined as the reciprocal of the energy ($E_{CH}$) consumed in information transmission when assuming the cluster head, and the income of the rest of the nodes that become cluster members is defined as the reciprocal of the energy ($E_{CM}$) consumed when completing their own information transmission.

This kind of income setting is in line with the actual situation, because $E_{CH} > E_{CM}$, cluster members are more profitable, and nodes prefer to be cluster members.

However, because all nodes want to become cluster members to maximize their survival time, this will lead to the lack of cluster heads in the network, which will lead to greater overall losses. Therefore, in order to avoid the situation that there is no cluster head in the network and the nodes suitable for acting as cluster heads do not actively act as cluster heads, we introduce a node punishment mechanism. Specifically, when the residual energy of a node is lagre, it will not consume energy quickly after acting as cluster head, so it is more suitable to act as cluster head. If the node is unwilling to act as cluster head, it will be punished to reduce its income. In addition, for nodes with a large number of neighboring nodes, they also need to act as cluster heads to fuse surrounding data to reduce network energy consumption. If such nodes do not act as cluster heads, they should also be punished. Therefore, the design of the node utility function is shown in (11).

$$U(s_i) = \begin{cases} 0, & S_i = CM, \quad \forall i \in N_{b_{S_i}}, \; S_j = CM \\ \dfrac{1}{E_{CH(S_i)}}, & S_i = CH \\ \dfrac{\Phi(S_i)}{E_{CM(S_i)}}, & S_i = CM, \quad \exists j \in N_{b_{S_i}}, \; S_j = CH \end{cases}$$
(11)

Here, $E_{CM(S_i)}$ and $E_{CH(S_i)}$ represent the cost of $S_i$ as a cluster member or a cluster head, respectively. $\Phi(S_i)$ denotes the penalty coefficient and $\Phi(S_i) \in [0, 1]$. The value of penalty factor is related to the residual energy and the neighbor number of $S_i$. When a node has more residual energy, compared with a node with low energy, it will not consume energy quickly when it becomes the cluster head, so if it refuses to become the cluster head, we will punish it. In addition, when the number of neighbors of a node is large, it is more necessary for the cluster head to complete the collection and forwarding tasks. If it refuses to become the cluster head, we will also punish it. Therefore, the expression of $\Phi(S_i)$ can be computed by (12).

$$\Phi(S_i) = \alpha \frac{E_{R_{max}} - E_{R_{S_i}}}{E_{R_{max}} - E_{R_{min}}} + \beta \frac{N_{b_{max}} - N_{b_{S_i}}}{N_{b_{max}} - N_{b_{min}}} \quad (12)$$

where, $\alpha$ and $\beta$ are the control parameters in the range [0,1], with $\alpha + \beta = 1$. $E_{R_{S_i}}$ and $N_{b_{S_i}}$ represent the residual energy and the number of neighboring nodes of $S_i$, respectively. $E_{R_{max}}$ and $E_{R_{min}}$ represent the maximum and minimum residual energy in neighbor nodes of $S_i$, respectively. $N_{b_{max}}$ and $N_{b_{min}}$ represent the maximum and minimum node degree (the number of neighbor nodes) in neighbor nodes of $S_i$, respectively.

Based on CROSS [12] and LGCA [14] protocols, in the clustering game, only when the nodes adopt a mixed strategy, there will be a symmetric Nash Equilibrium. That is, the node becomes the cluster head with probability $p$ and the cluster member with probability $q = 1 - p$. According to the definition of mixed Nash equilibrium [19], the expected utilities of playing strategies $CH$ and $CM$ are equal and no player has incentive to change her strategy. Thus, there are:

$$U_{CH} = U_{CM}. \tag{13}$$

It can be deduced from (12) that the income $U_{CH}$ when the node becomes the cluster head and the income $U_{CM}$ of the cluster member are shown in (14) and (15) respectively.

$$U_{CH} = \frac{1}{E_{CH}} \tag{14}$$

$$\begin{aligned} U_{CM} &= Pr\{S_i = CM, \forall j \in N_{b_{S_i}}, S_j = CM\} \cdot 0 + \cdots \\ &+ Pr\{S_i = CM, \exists j \in N_{b_{S_i}}, S_j = CH\} \cdot \frac{\Phi}{E_{CM}} \\ &= \frac{\Phi}{E_{CM}} \cdot (1 - Pr\{S_i = CM, \forall i \in N_{b_{S_i}}\}) \\ &= \frac{\Phi}{E_{CM}} \cdot [1 - (1-p)^{\frac{1}{N_{b_{S_i}}-1}})] \end{aligned} \tag{15}$$

Substituting (14) and (15) in (13) and solving the expression in order to calculate the probability $p$ that corresponds to the equilibrium, we can get the probability $p_i$ of becoming the cluster head of a node, as shown in (16).

$$p_i = 1 - (\frac{\Phi E_{CH} - E_{CM}}{\Phi E_{CH}})^{\frac{1}{N_{b_{S_i}}-1}} \tag{16}$$

### B. Cluster Head Selection and Cluster Formation

*1) Selection of Candidate CH:* In this sub phase, each nodes will play several clustering games with its neighbor nodes. After the ending of the clustering game, each node $S_i$ will calculate the equilibrium probability according to (16) and generate the random number $rand_i \in [0, 1]$, so as to decide itself whether to be a candidate cluster head. To be specific, for node $S_i$, if $p_i \geqslant rand_i$, node $S_i$ will be selected the candidate cluster head. At this time, all candidate cluster heads will form candidate cluster head set $C$.

*2) Final Cluster Head Selection:* Since more than one candidate cluster head selected in the previous sub-phase may occur in a certain region, which leads to the nonuniform distribution of cluster heads. Worse, the number of the selected candidate cluster head deviates from the optimal cluster head number ($K_{opt}$), which is not conducive to network energy efficiency. Therefore, we propose a coverage optimization based final cluster head selection algorithm to overcome the aforementioned problems. The principle of this cluster head selection algorithm is as follows.

According to the definition of coverage [20], [21], we first define coverage value $\Gamma$ according whether a node is covered by the cluster head, as shown in (17). That is, if the distance between cluster head $CH_k$ and node $S_j$ does not exceed $R$, the node is considered to be covered by cluster head, and the coverage value $\Gamma(CH_k, S_j)$ is 1. Otherwise, the coverage value is 0. As shown in Fig. 2, $S_1$ is covered by $CH_1$, so $\Gamma(CH_1, S_1)$ is 1, conversely, $\Gamma(CH_1, S_2)$ is 1 because $S_1$ is not located in $CH_1$'s coverage range $R$.

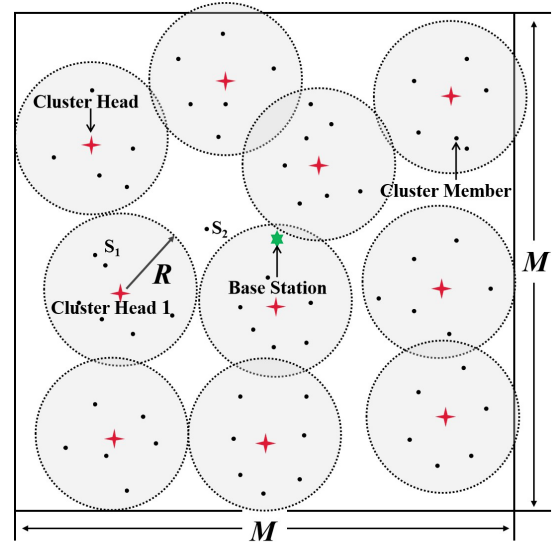

Fig. 2. Multi hop communication between clusters.

$$\Gamma(CH_k, S_j) = \begin{cases} 0, & d(CH_k, S_j) > R \\ 1, & otherwise. \end{cases} \tag{17}$$

As node $S_j$ may be covered by multiple cluster heads at the same time, the coverage value of a cluster head covering set to node $S_j$ is defined as (18).

$$\Gamma(CH_k, S_j) = 1 - \prod_{i=1}^{N}(1 - \Gamma(CH_i, S_j)) \tag{18}$$

Then, the coverage rate of a cluster head coverage set to all nodes in the region $\Gamma_{Total}$ can be defined as the ratio of the coverage value of all sensor nodes covered by the cluster head covering set to the total number of nodes $N$ in the region, as shown in (19).

$$\Gamma_{total} = \frac{\sum_{j \in N} \Gamma(CH, S_j)}{N} \tag{19}$$

As depicted in Fig. 2, when all nodes are covered by cluster heads, that is, when the coverage rate $\Gamma_{Total}$ is as close as possible to 1, cluster heads will be evenly distributed, which will optimize the communication distance between cluster heads and cluster members. However, residual energy is another important factor in choosing cluster head. If a low-energy node is selected as a cluster head, it will die quickly and shorten the network lifetime. Therefore, In order to give consideration to the coverage uniformity and residual energy, we set the objective function related to coverage and residual energy, as shown in (2).

$$F = \lambda_1 \frac{\sum_{j \in N} \Gamma(CH, S_j)}{N} + \lambda_2 \frac{\sum_{CH_k \in C} E_R(CH_k)}{E_O \times N} \tag{20}$$

Here, $\lambda_1$ and $\lambda_2$ represent coverage factor and residual energy factor, respectively. $E_R$ and $E_O$ represent the residual energy and initial energy of candidate cluster head, respectively. $\mathcal{C}$ is the candidate cluster head set.

Particle swarm optimization (PSO) [22], as a swarm intelligence optimization algorithm, has the advantages of simple implementation and high search accuracy, and has great advantages in solving combinatorial optimization problems [23] such as optimal cluster head covering set. Therefore, this paper uses bat algorithm to find the optimal solution of objective function $F$. After the PSO algorithm iteration, the output optimal cluster head node set may not really exist in the monitoring area. At this time, a location mapping strategy is needed. The basic idea is to find out the candidate cluster head in the actual network which is closest to the optimal solution output by PSO algorithm and become the final cluster head. After the final cluster head is determined, the node that becomes the cluster head will inform other nodes by broadcasting, and the rest nodes will join according to the cluster head with the highest received signal strength to form a cluster.

## C. Data Transmission

For large-scale WSNs, if the relay strategy is not used, there will be a large number of cluster heads that communicate with the base station in the multipath fading model far away from the base station, resulting in great network overhead. However, in the existing multi-hop routing protocols, the remaining energy of the relay node, the distance from the base station, the load and other factors are often considered when selecting the relay node, and it is impossible to mathematically deduce and analyze whether the relay node can save energy [24], [25]. Therefore, in order to construct relay forwarding strategies between cluster heads in different scenarios and transmit intra-cluster data to base stations with lower energy consumption, **Theorem 1** is proposed as follows.

*Theorem 1:* For WSNs based on clustering routing protocol, the relay strategy can reduce the network energy consumption when the distance between the sending node and the relay node ($d_{toRe}$) meets (21) whose limiting condition are shown in (22). $d_{toDest}$ represents the distance from source node to de the stination node. $d_{RtoD}$ represents the distance from the relay node to the destination node.

$$\begin{cases} d_{toRe} < (d_{toDest}^2 - \dfrac{N_C(\varepsilon_{elec} + \varepsilon_{fs}d_{RtoD}^2)}{\varepsilon_{fs}})^{\frac{1}{2}} \\ d_{toRe} < (\dfrac{\varepsilon_{mp}d_{toDest}^4 - N_C(\varepsilon_{elec} + \varepsilon_{fs}d_{RtoD}^2)}{\varepsilon_{fs}})^{\frac{1}{2}} \\ d_{toRe} < (d_{toDest}^2 - \dfrac{N_C(\varepsilon_{elec} + \varepsilon_{mp}d_{RtoD}^4)}{\varepsilon_{fs}})^{\frac{1}{2}} \\ d_{toRe} < (\dfrac{\varepsilon_{mp}d_{toDest}^4 - N_C(\varepsilon_{elec} + \varepsilon_{mp}d_{RtoD}^4)}{\varepsilon_{fs}})^{\frac{1}{2}} \\ d_{toRe} < (\varepsilon_{fs}d_{toDest}^2 - \dfrac{N_C(\varepsilon_{elec} + \varepsilon_{fs}d_{RtoD}^2)}{\varepsilon_{mp}})^{\frac{1}{4}} \\ d_{toRe} < (d_{toDest}^4 - \dfrac{N_C(\varepsilon_{elec} + \varepsilon_{fs}d_{RtoD}^2)}{\varepsilon_{mp}})^{\frac{1}{4}} \\ d_{toRe} < (\varepsilon_{fs}d_{toDest}^2 - \dfrac{N_C(\varepsilon_{elec} + \varepsilon_{mp}d_{RtoD}^4)}{\varepsilon_{mp}})^{\frac{1}{4}} \\ d_{toRe} < (d_{toDest}^4 - \dfrac{N_C(\varepsilon_{elec} + \varepsilon_{mp}d_{RtoD}^4)}{\varepsilon_{mp}})^{\frac{1}{4}} \end{cases} \quad (21)$$

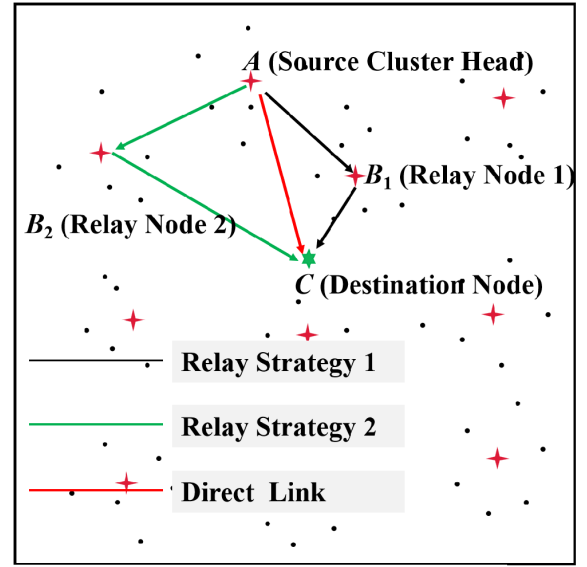

Fig. 3. Multi hop communication between clusters.

$$\begin{cases} d_{toRe} \leq d_o, & d_{RtoD} \leq d_o, & d_{toDest} \leq d_o \\ d_{toRe} \leq d_o, & d_{RtoD} \leq d_o, & d_{toDest} > d_o \\ d_{toRe} \leq d_o, & d_{RtoD} > d_o, & d_{toDest} \leq d_o \\ d_{toRe} \leq d_o, & d_{RtoD} > d_o, & d_{toDest} > d_o \\ d_{toRe} > d_o, & d_{RtoD} \leq d_o, & d_{toDest} \leq d_o \\ d_{toRe} > d_o, & d_{RtoD} \leq d_o, & d_{toDest} > d_o \\ d_{toRe} > d_o, & d_{RtoD} > d_o, & d_{toDest} \leq d_o \\ d_{toRe} > d_o, & d_{RtoD} > d_o, & d_{toDest} > d_o \end{cases} \quad (22)$$

*Proof:* According to the structure of clustering routing protocol, multi-hop relay strategy only affects the communication between cluster head and base station, and has no effect on the communication energy consumption between member nodes and cluster head. Therefore, when analyzing the performance of relay strategy, we only discuss the energy consumption of cluster head. The energy consumption of the relay cluster head includes receiving the sensing data from the cluster head and the cluster head to be relayed, merging the data in the cluster and forwarding the data sent by the cluster head to the destination node (cluster head or base station), as shown in (23).

$$E_{CR} = \begin{cases} l\varepsilon_{elec}(N_{CM} - 1 + N_C) + l\varepsilon_{DA}N_{CM} \\ + (1 + N_C)l(\varepsilon_{elec} + \varepsilon_{fs}d^2) & d \leq d_o \\ l\varepsilon_{elec}(N_{CM} - 1 + N_C) + l\varepsilon_{DA}N_{CM} \\ + (1 + N_C)l(\varepsilon_{elec} + \varepsilon_{mp}d^4) & d > d_o \end{cases} \quad (23)$$

Now, let's compare the energy consumption difference before and after a relay. As depicted in Fig. 3, source node $A$ can directly transmit data to destination node $C$, or it can transmit data to $C$ through relay node $B_1$. For the sake of theoretical deduction, we define that the total energy consumption of node $A$ transmitting data directly to node $C$ is marked as $E_{NR}$ and the total energy consumption of node $A$ transmitting data to node $C$ through relay node $B_1$ is marked as $E_R$. Therefore, $E_R - E_{NR}$ is the energy saved by relay, as shown in (24) with the limiting conditions revealed by (22). For large-scale WSNs, the communication between nodes can be divided into free space model and multipath fading model, so (24) includes eight cases.

$$E_{NR} - E_R = \begin{cases} \varepsilon_{fs}(d_{toDest}^2 - d_{toRe}^2) + N_C(\varepsilon_{elec} + \varepsilon_{fs}d_{RtoD}^2) \\ \varepsilon_{fs}d_{toDest}^2 - \varepsilon_{mp}d_{toRe}^4 + N_C(\varepsilon_{elec} + \varepsilon_{fs}d_{RtoD}^2) \\ \varepsilon_{fs}(d_{toDest}^2 - d_{toRe}^2) + N_C(\varepsilon_{elec} + \varepsilon_{mp}d_{RtoD}^4) \\ \varepsilon_{fs}d_{toDest}^2 - \varepsilon_{mp}d_{toRe}^4 + N_C(\varepsilon_{elec} + \varepsilon_{mp}d_{RtoD}^4) \\ \varepsilon_{mp}d_{toDest}^4 - \varepsilon_{fs}d_{toRe}^2 + N_C(\varepsilon_{elec} + \varepsilon_{fs}d_{RtoD}^2) \\ \varepsilon_{mp}(d_{toDest}^4 - d_{toRe}^4) + N_C(\varepsilon_{elec} + \varepsilon_{fs}d_{RtoD}^2) \\ \varepsilon_{mp}d_{toDest}^4 - \varepsilon_{fs}d_{toRe}^2 + N_C(\varepsilon_{elec} + \varepsilon_{mp}d_{RtoD}^4) \\ \varepsilon_{mp}(d_{toDest}^4 - d_{toRe}^4) + N_C(\varepsilon_{elec} + \varepsilon_{mp}d_{RtoD}^4) \end{cases} \quad (24)$$

When the value of (24) is greater than 0, compared with the communication mode of one-hop transmission to the base station, the energy consumption of relay multi-hop is reduced, and the role of relay nodes can be reflected. The solution (21) at this time is the applicable range of relay forwarding strategy. The proof is completed.

Based on **Theorem 1**, we propose a relay node selection strategy. The cluster head nodes communicate with each other and store the residual energy and location information, and judge the candidate relay nodes that meet the above-mentioned multi-hop applicable conditions, and using (25) select the cluster head with large residual energy and close to the base station as the final relay node among its candidate relay nodes.

$$F(CH_j) = \frac{E_R(CH_j)}{N_b(CH_j) \times \|BS, CH_j\|} \quad (25)$$

Here, $E_R$ represents the residual energy of relay cluster head $CH_j$. $N_b$ is the neighbor node number of cluster head $CH_j$. $\|BS, CH_j\|$ is euclidean distance between cluster head $CH_j$ and base station.

### D. Computational Complexity Analysis

For a network with $N$ nodes and $K$ cluster heads and $N \gg K$, the computational complexity of the candidate cluster head selection based on clustering game is $O(N)$; The computational complexity of the final cluster head selection based on PSO is $O(N_{iter} \times N_{pop} \times N)$, and the final cluster formation complexity is $O(N \times K)$. Here, $N_{iter}$ and $N_{pop}$ represent the number of iteration times and population number, respectively. During the communication between the cluster head and the base station, the computational complexity of relay node selection is $O(K^2)$. Considering that the number of cluster heads ($K$) is much smaller than the product of population number and PSO iterations, the computational complexity of MRP-GTCO proposed in this paper is $O(N_{iter} \times N_{pop} \times N)$ during each round of cluster head election and communication.

## IV. EXPERIMENTAL SIMULATION AND ANALYSIS

This section provides the experimental results to evaluate the proposed MRP-GTCO protocol's efficiency for prolonging network lifetime and reducing network energy consumption. Firstly, we discussed the parameter setting of MRP-GTCO protocol. Then, some performance indicators such as cluster head number, cluster head energy consumption, network lifetime and network energy consumption are compared with RLEACH [26], LGCA ($w = 0.05$ and $R_c = 30m$), and ECAGT ($w = 0.05$, $R_c = 30m$ and $\alpha = 8$) protocols. All the routing protocols were coded and executed on MATLAB R2016b.

*Remark 2:* Note that in order to better understand the relationship between network lifetime and network energy consumption, we conclude that network energy consumption is the most important factor that affects HDN and LDN. In the case of the same total network energy, if the network's energy consumption in each round is lower, HDN and LDN will be longer. Cluster heads take on more forwarding tasks and consume more energy than cluster members, so FDN is mainly affected by the death round of cluster heads.

### A. Verification of Optimal Cluster Head Number

In this subpart, taking the $200m \times 200m$ monitoring area with $N = 100$ as an example, we discuss the changing trend of network energy consumption with the number of cluster heads and provide the optimal parameters for subsequent simulation. The values of energy consumption coefficient of wireless transceiver circuit $E_{elec}$, data fusion energy consumption coefficient $E_{DA}$ and free space energy consumption coefficient $E_{fs}$ are set to $50nJ/bit$, $5nJ/bit$ and $10pJ/bit/m^2$, respectively. Note that presumably the distance between the cluster head and its relay node ranges form $0m$ to $141m$ for $200m \times 200m$ monitoring area. Secondly, it is assumed that the number of relay data forwarding times of each cluster head is about $2 - 5$ times, namely $2 \leq N_C \leq 5$. Thus, based on (9), the optimal number of cluster heads can be calculated ranging from 4 to 11 nodes. Lastly, by changing the number of cluster heads in the range of 2 to 14, the average network energy consumption before the first dead node appears is counted to verify the optimal cluster head number.

In Fig. 4, the trend of network energy consumption with the number of cluster heads is illustrated. We can draw a conclusion that with the increase of cluster heads, the network energy consumption first increases and then decreases, and when the number of cluster heads is 10 nodes, the network energy consumption reaches the minimum, which is well matched with the theoretical analysis value of $K_{opt}$. Therefore, we will set $K_{opt}$ to 10 in $200m \times 200m$ monitoring area with $N = 100$ nodes for the subsequent experiments.

### B. Influence of Energy and Coverage Weight on Network

The energy consumption of cluster head is generally higher than that of cluster members, because it undertakes the huge energy consumption task of data fusion and forwarding of nodes in the cluster. After the candidate cluster heads are determined, the final cluster head selection of the MRP-GTCO protocol is to consider both coverage and residual energy using (20) where $\lambda_1$ and $\lambda_2$ control the weight of coverage and residual energy factor respectively. So how to select them optimally is the key to extend network lifetime and reduce energy consumption. Therefore, we change $\lambda_1$ from 0 to 1 in step of 0.1 to analyze the variations of network energy consumption and network lifetime, as demonstrated in Fig. 5 and Fig. 6, respectively.

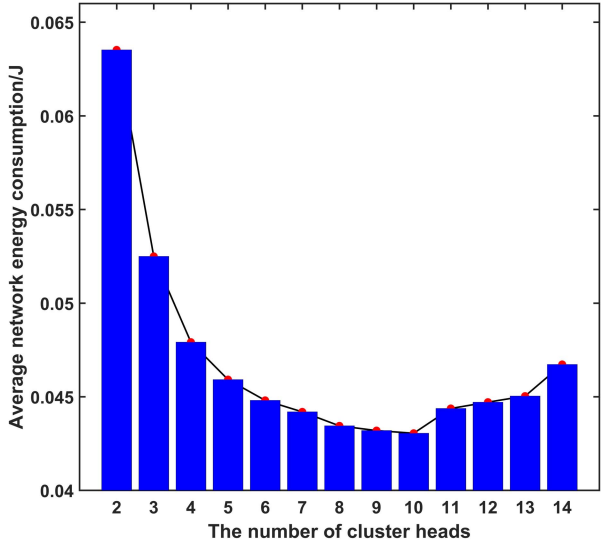

Fig. 4. Relationship between network energy consumption and number of cluster heads ($200m \times 200m$ monitoring area with $N = 100$ nodes).

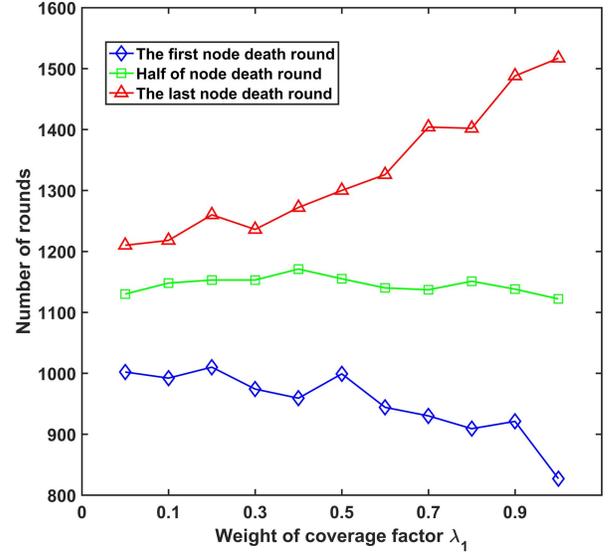

Fig. 5. Relationship between network lifetime and coverage factor weight $\lambda_1$ ($200m \times 200m$ monitoring area with $N = 100$ nodes).

TABLE III
EXPERIMENTAL PARAMETERS

| Parameters | Values |
|---|---|
| Monitoring area size $M \times M$ | $200m \times 200m$ |
| | $300m \times 300m$ |
| | $400m \times 400m$ |
| $K_{opt}$ | 10 nodes; 12 nodes; 15 nodes |
| Number of nodes $N$ | 100 nodes |
| Node initial energy $E_o$ | $0.5\ J$ |
| Data packet size $l$ | $4000\ bits$ |
| $E_{mp}$ | $0.0013\ pJ/bit/m^4$ |
| $E_{fs}$ | $10\ pJ/bit/m^2$ |
| $E_{elec}$ | $50\ nJ/bit$ |
| $E_{DA}$ | $5\ nJ/bit$ |
| Coverage factor weight $\lambda_1$ | 0.5 |
| Energy factor weight $\lambda_2$ | 0.5 |
| Energy factor $\alpha$, node degree factor $\beta$ | 0.7, 0.3 |

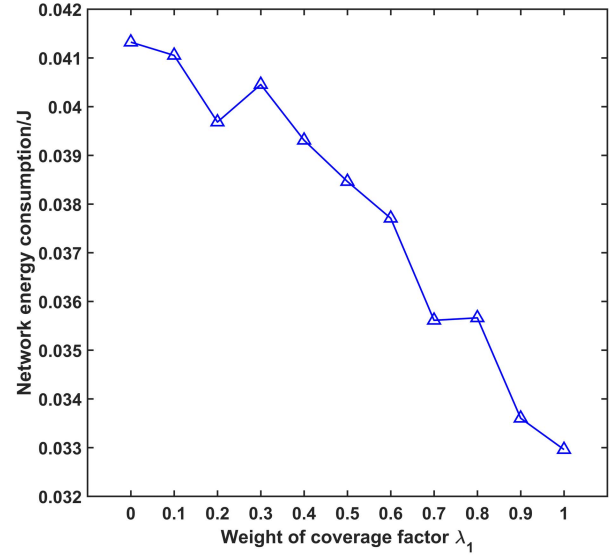

Fig. 6. Relationship between network energy consumption and coverage factor weight $\lambda_1$ ($200m \times 200m$ monitoring area with $N = 100$ nodes).

Fig. 5 clearly shows that With the increase of $\lambda_1$, the first death node round (FDN) showed a downward trend, half of death nodes round (HDN) remained basically unchanged, while the last death node round (LDN) kept an upward trend. Furthermore, in Fig. 6, the network energy consumption decreases with the increase of $\lambda_1$. The reasons for the above phenomena can be summarized as follows. With the increase of the coverage factor weight $\lambda_1$, the selected cluster heads tend to ensure the coverage of cluster heads and nodes (i.e., uniformity of cluster head distribution). The more uniform distribution of cluster heads can effectively reduce the communication distance between cluster heads and other nodes (cluster members or relay nodes), thus reducing the network energy consumption. Therefore, it is not difficult to understand that the lower the network energy consumption, the death rounds of all nodes can be postponed. However, with the increase of the coverage factor weight $\lambda_1$ (i.e., the decrease of the energy factor weight $\lambda_2$), the selected cluster head will have less residual energy, which makes the selected cluster head die quickly, so FDN will decrease continuously. Since FDN is the most important representation of network lifetime, at the same time, giving rise to an interesting trade-off between FDN and LDN, we will choose $\lambda_1 = 0.5$ for follow-up experiments.

The parameters involved in the following comparative experiments are shown in Table III.

*1) Experiments Based on Cluster Heads:* According to the experimental analysis in Section IV. part B, when the number of cluster heads is kept at about 10, the network energy consumption can be minimized. Therefore, it is important to compare the change of cluster head number of four protocols before the death of the first node. As can be seen from Fig. 7, The number of cluster heads of the four protocols fluctuates to varying degrees with the network operation. The number of cluster heads of the RLEACH protocol fluctuates between 2 and 26, and the fluctuation range is the largest. The number of cluster heads of LGCA and ECAGT protocols keeps about range from 9 to 23 and fluctuates. However, the number of cluster heads of the proposed MRP-GTCO protocol is maintained at about from 6 to 10, and most cases can fluctuate around the optimal number of cluster heads (10 nodes), which will be conducive to ensuring that MRP-GTCO protocol can run with low network energy consumption and ensure energy efficiency.

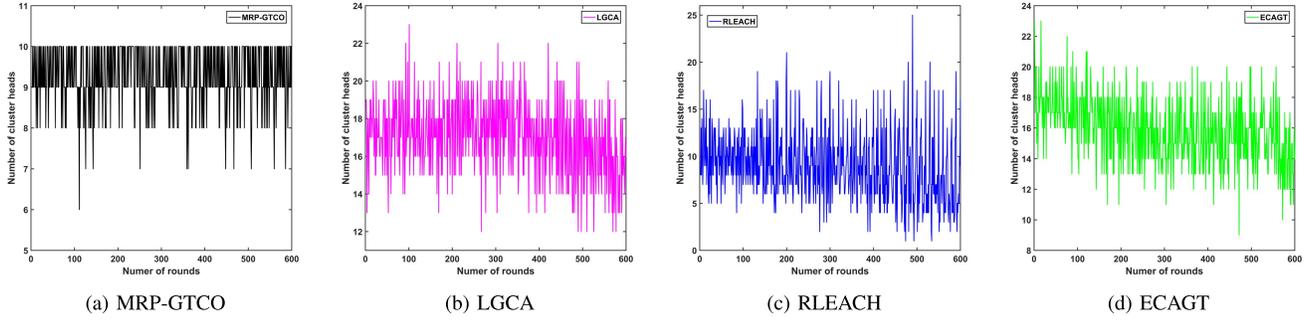

Fig. 7. The number of cluster heads (200*m* × 200*m* monitoring area with *N* = 100 nodes).

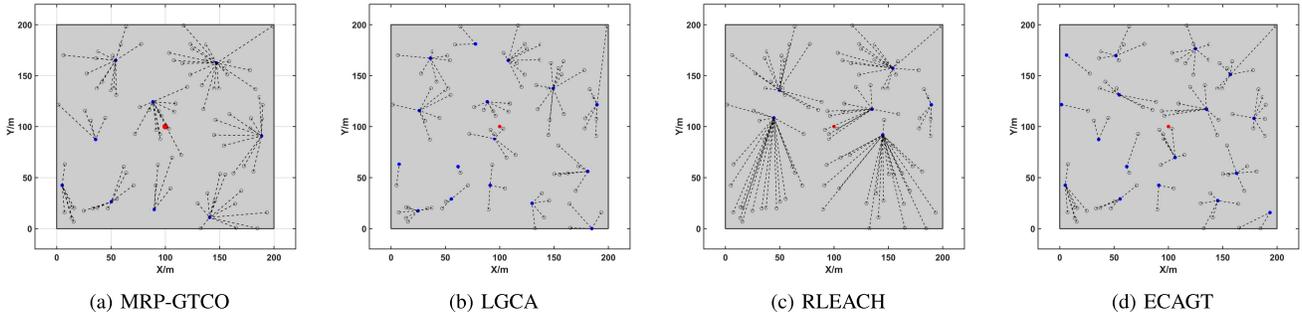

Fig. 8. Comparison of cluster head distribution. The red and blue solid dots represent the base station and cluster head, respectively. Black circles represent cluster member nodes. (200*m* × 200*m* monitoring area with *N* = 100 nodes.)

The uniformity of cluster head distribution is also an important factor to reduce network energy consumption. The more uniform the distribution of cluster heads, the more reasonable the distance between cluster heads and cluster member nodes or relay cluster heads, thus reducing the communication energy consumption between nodes. Fig. 8 shows the cluster head distribution of MRP-GTCO, LGCA, RLEACH and ECAGT protocols. We can deduce that the cluster heads of MRP-GTCO are uniformly distributed in the detection area, and the distance between cluster heads and cluster members or other cluster heads is reasonable, and the number of cluster heads is 9, which is close to the optimal number of cluster heads. For the other three protocols, the number of cluster heads obviously deviates from the optimal number and distribution of cluster heads is nonuniform.

Since cluster heads consume more energy and are more likely to die, the cluster head energy consumption plays a

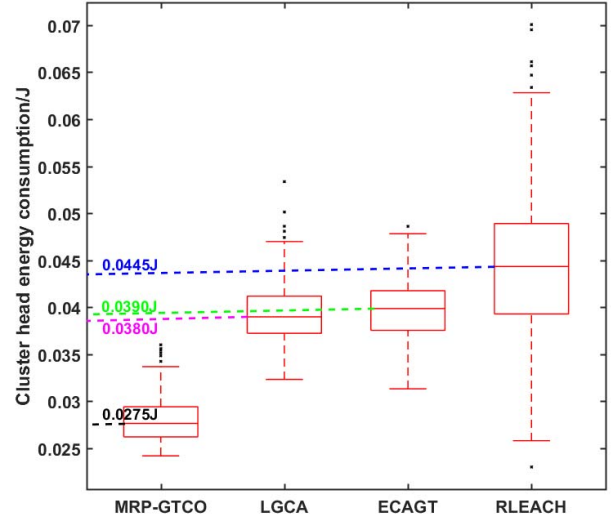

Fig. 9. The boxplot of cluster head energy consumption of four protocols in first 600 rounds (200*m* × 200*m* monitoring area with *N* = 100 nodes).

decisive role in the first dead node round (FDN) of the network. Fig. 9 gives the comparison of cluster head energy consumption among MRP-GTCO, LGCA, ECAGT and RLEACH. In Fig. 9, we can summarize that the middle bit line, upper bit line and lower bit line of MRP-GTCO protocol are lower than RLEACH, LGCA and ECAGT protocols, and the fluctuation between upper and lower bit lines of MRP-GTCO protocol is also the smallest. Especially, the median energy consumption of EERP is 38.19%, 41.82% and 61.81% lower than that of

LGCA, ECAGT and RLEACH respectively. Therefore, the proposed MRP-GTCO protocol can effectively reduce the cluster head energy consumption, which will help to prolong FDN.

*2) Experiments Based on Network:* Network energy consumption is defined as the sum of energy consumption of all nodes in the network operation, which can be reflected by the change of network residual energy with the network operation rounds. As can be seen from Fig. 10, MRP-GTCO protocol maintains a high residual energy before and after the stable period of network operation (about 1100th rounds). Especially in the 800th rounds, the residual energy of MRP-GTCO, ECAGT, LGCA and RLACH protocols is 15.63J, 13.50J, 13.09J and 4.29J respectively. Compared with ECAGT, LGCA and RLACH protocols, MRP-GTCO can save 15.78%, 19.41% and 264.34% of energy respectively. However, after the 1000th rounds, the residual energy of MRP-GTCO protocol network is slightly lower than that of LGCA. The main reason is that the number of surviving nodes of MRP-GTCO is less after the 1000th rounds, which makes it difficult to optimally match the real cluster head position mapped by the final cluster first selection algorithm with the virtual cluster head selected by PSO, thus leading to the poor cluster head being selected to undertake the network forwarding task and consuming a lot of energy, resulting in less network residual energy.

Network lifetime is the most important index to measure the energy efficiency of the algorithm, which is importantly characterized by the first dead node round (FDN) because before the first dead node appeared, the network topology did not change and the collected information was the most comprehensive. Fig. 11 reflects the variation of the number of surviving nodes of the four protocols with the network running rounds. It is not difficult to see from Fig. 11 that the FDN of MRP-GTCO, ECAGT, LGCA and RLACH protocols are about 998, 857, 385 and 662, respectively. The MRP-GTCO protocols extend FDN by 16.46%, 159.22% and 50.76% compare to ECAGT, LGCA and RLEACH protocols, respectively. In addition, for MRP-GTCO protocol, the difference between FDN and LDN (Last dead node) is the smallest, which demonstrates that all nodes of MRP-GTCO protocol die almost at the same time and the energy consumption of each node keeps almost the same level.

### C. Influence of Network Area on MRP-GTCO

In this section, by setting the size of the monitoring area to $300m \times 300m$ and $400m \times 400m$, the FDN, HDN (Half dead node), LDN and energy utilization rate of MRP-GTCO-noRelay, MRP-GTCO, ECAGT, LGCA and RLEACH protocols in the two scenarios are counted respectively, and finally the influence of the area expansion on the five protocols and the effectiveness of the proposed multi-hop path selection algorithm in this paper are analyzed. Note that here, MRP-GTCO-noRelay represents MRP-GTCO protocol that does not use the multi-hop routing algorithm proposed in this paper.

*1) Energy utilization efficiency:* Network energy utilization rate is an important evaluation index to measure network energy efficiency, which can be reflected by the change of the number of alive nodes and packets received by the base station with the consumption of network energy.

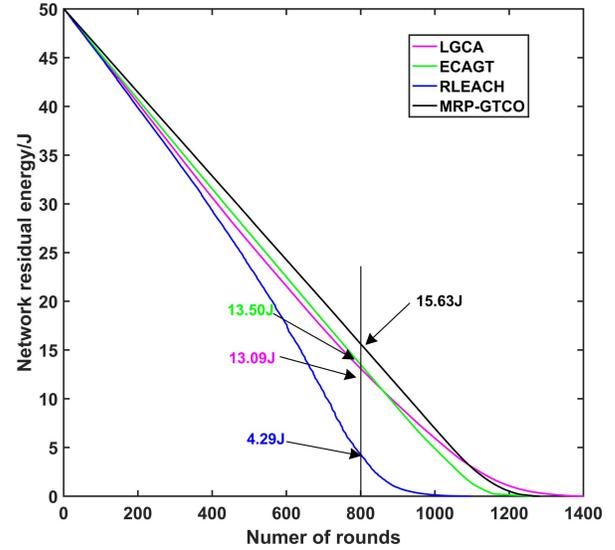

Fig. 10. Comparison of network residual energy ($200m \times 200m$ monitoring area with $N = 100$ nodes).

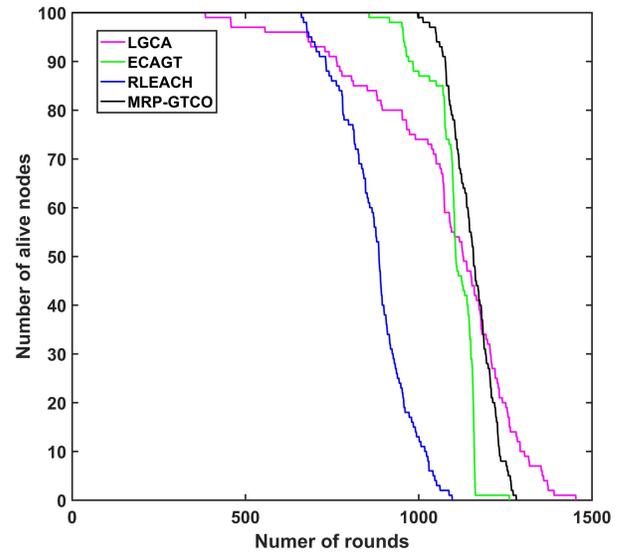

Fig. 11. Comparison of alive node number change ($200m \times 200m$ monitoring area with $N = 100$ nodes).

Fig. 12(a) and Fig. 12(b) show the energy utilization rate of MRP-GTCO- noRelay, MRP-GTCO, ECAGT, LGCA and RLEACH protocols under $300m \times 300m$ and $400m \times 400m$ scenarios, respectively. In Fig. 12(a), when the network consumes $40J$ energy, the number of surviving nodes of the MRP-GTCO, MRP-GTCO-noRelay, ECAGT, LGCA and RLEACH protocols is about 98, 92, 78, 72 and 60 respectively; In the $400m \times 400m$ scenario, the number of surviving nodes of the MRP-GTCO, MRP-GTCO- noRelay, ECAGT, LGCA and RLEACH protocols is about 85, 73, 68, 56 and 56 respectively. Therefore, no matter what scenario MRP-GTCO consumes the same energy, it has the most alive nodes. Therefore, MRP-GTCO has the highest energy utilization rate.

In addition, Fig. 12(c) and Fig. 12(d) depict the energy utilization rate of MRP-GTCO- noRelay, MRP-GTCO, ECAGT,

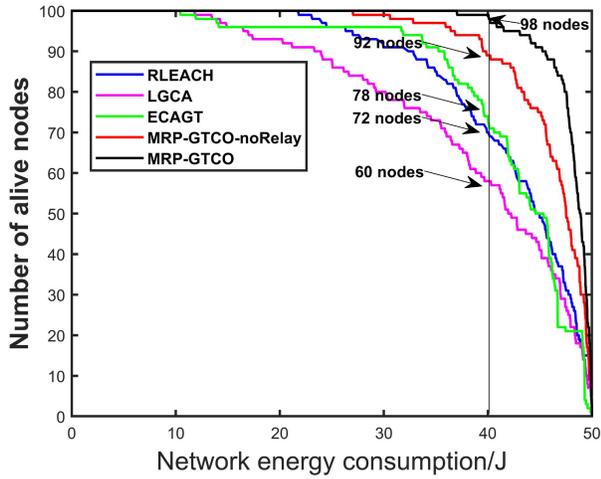
(a) $300m \times 300m$ monitoring area with $N = 100$ nodes

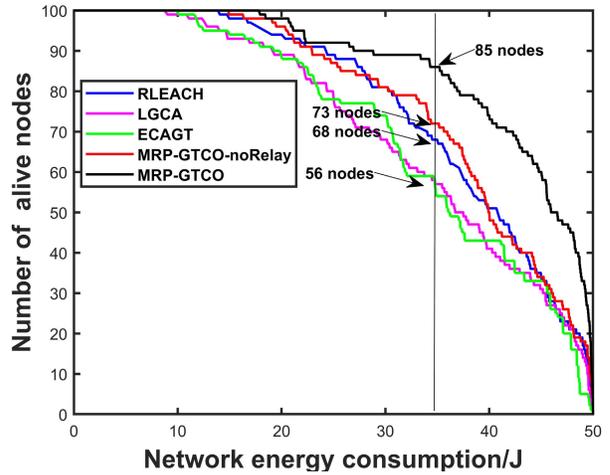
(b) $400m \times 400m$ monitoring area with $N = 100$ nodes

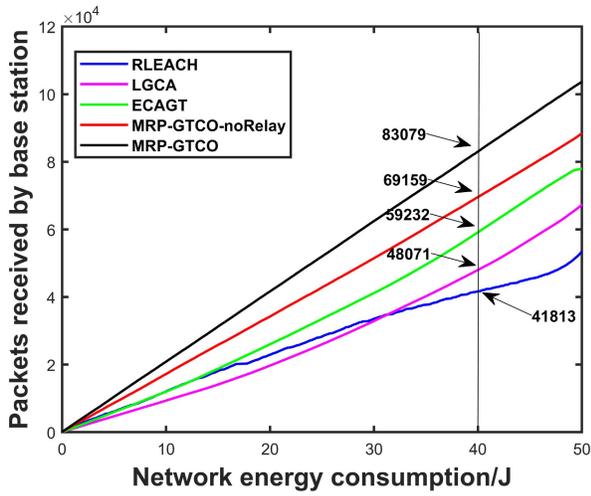
(c) $300m \times 300m$ monitoring area with $N = 100$ nodes

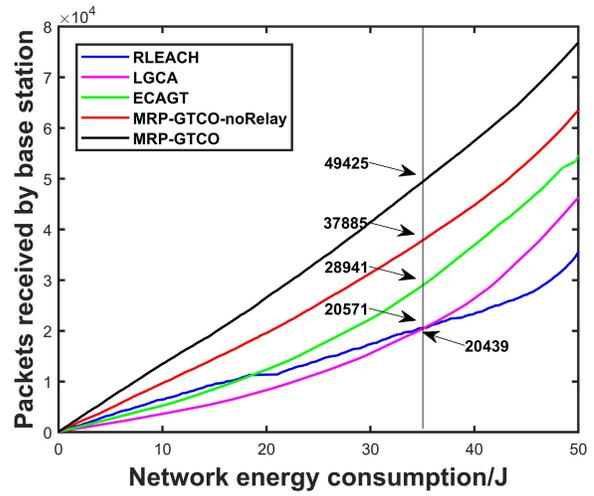
(d) $400m \times 400m$ monitoring area with $N = 100$ nodes

Fig. 12. Energy utilization efficiency.

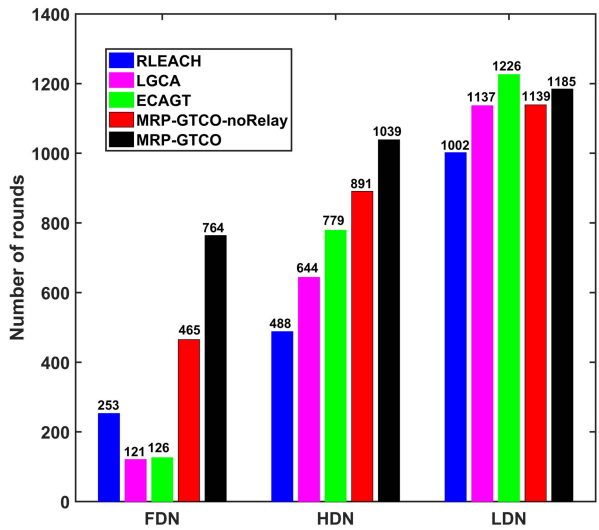
(a) $300m \times 300m$ monitoring area with $N = 100$ nodes

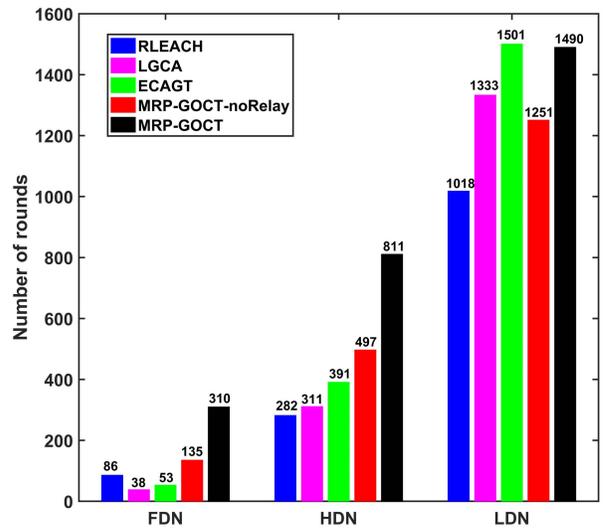
(b) $400m \times 400m$ monitoring area with $N = 100$ nodes

Fig. 13. Comparison of network lifetime in different scenarios.

LGCA and RLEACH protocols on the number of packets received by base station under $300m \times 300m$ and $400m \times 400m$ scenarios, respectively. We can conclude that when the network consumes the same energy, MRP-GTCO can always transmit more data packets to the base station. Hence, the high energy efficiency of MRP-GTCO is reflected.

*2) Network lifetime:* Fig. 13 shows the comparison chart of network lifetime in two scenarios. Fig. 13(a) shows the simulation results in the scene of $300m \times 300m$ where the optimal number of cluster heads for MRP-GTCO-noRelay and MRP-GTCO protocol is set to 12, and it can be seen that FDN and HDN of MRP-GTCO protocol are outperform the other three protocols. Fig. 13(b) shows the simulation results in the scene of $400m \times 400m$ where the optimal number of cluster heads for MRP-GTCO-noRelay and MRP-GTCO protocol is set to 15, and it can be seen that FDN and HDN of MRP-GTCO protocol are obviously superior to the LGCA, ECAGT and RLEACH protocols. By comparing Fig. 13(a) and Fig. 13(b), we can cout that with the increase of regional scale, the FDN of MRP-GTCO protocol decreases by 59.42%, and the other three protocols decrease by 66.1%, 72.73% and 72.16% respectively, which shows that MRP-GTCO protocol has better adaptability with the change of network scale. In addition, in two monitoring scenarios, it can be clearly seen that FDN, LDN and HDN of MRP-GTCO protocol are higher than those of MRP-GTCO-noRelay protocol, which verifies that the multi-hop path selection algorithm proposed in this paper can effectively reduce the communication energy consumption among cluster heads, thus reducing the network energy consumption and optimizing the network lifetime.

## V. Conclusion

How to optimize the selection of cluster heads and relay path is the key to constructing high-performance multi-hop routing protocol. This paper presents a multi-hop routing protocol based on game theory and coverage optimization (MRP-GTCO), analyzes the applicable conditions of multi-hop strategy between clusters, reveals the mapping relationship between the number of cluster heads and network energy consumption, and deduces the optimal number of cluster heads in different scenarios. The optimal number of cluster heads and the effectiveness of multi-hop strategy are verified and analyzed by experiments. The experimental results show that, compared with LGCA, RLEACH and ECAGT, the MRP-GTCO protocol proposed in this paper effectively reduces the network energy consumption and prolongs the network lifetime. However, the last node death rounds of MRP-GTCO need to be further optimized, which will be considered in our future work. In addition, since the problems of communication congestion and data transmission rate caused by clustering and multi-hop between clusters have not been considered, the communication service quality will be researched in the future.

## References

[1] X. Zhao, Y. Cui, C. Gao, Z. Guo, and Q. Gao, "Energy-efficient coverage enhancement strategy for 3-D wireless sensor networks based on a vampire bat optimizer," *IEEE Internet Things J.*, vol. 7, no. 1, pp. 325–338, Jan. 2020.

[2] M. Torky and A. E. Hassanein, "Integrating blockchain and the Internet of Things in precision agriculture: Analysis, opportunities, and challenges," *Comput. Electron. Agricult.*, vol. 178, Nov. 2020, Art. no. 105476.

[3] T. Shu, J. Chen, V. K. Bhargava, and C. W. de Silva, "An energy-efficient dual prediction scheme using LMS filter and LSTM in wireless sensor networks for environment monitoring," *IEEE Internet Things J.*, vol. 6, no. 4, pp. 6736–6747, Aug. 2019.

[4] C. V. Mahamuni and Z. M. Jalauddin, "Intrusion monitoring in military surveillance applications using wireless sensor networks (WSNs) with deep learning for multiple object detection and tracking," in *Proc. Int. Conf. Control, Autom., Power Signal Process. (CAPS)*, Dec. 2021, pp. 1–6.

[5] P. Maheshwari, A. K. Sharma, and K. Verma, "Energy efficient cluster based routing protocol for WSN using butterfly optimization algorithm and ant colony optimization," *Ad Hoc Netw.*, vol. 110, Jan. 2021, Art. no. 102317.

[6] Y.-D. Yao, X. Li, Y.-P. Cui, J.-J. Wang, and C. Wang, "Energy-efficient routing protocol based on multi-threshold segmentation in wireless sensors networks for precision agriculture," *IEEE Sensors J.*, vol. 22, no. 7, pp. 6216–6231, Apr. 2022.

[7] Y. Yun, Y. Xia, B. Behdani, and J. C. Smith, "Distributed algorithm for lifetime maximization in a delay-tolerant wireless sensor network with a mobile sink," *IEEE Trans. Mobile Comput.*, vol. 12, no. 10, pp. 1920–1930, Oct. 2013.

[8] H. Yetgin, K. T. K. Cheung, M. El-Hajjar, and L. H. Hanzo, "A survey of network lifetime maximization techniques in wireless sensor networks," *IEEE Commun. Surveys Tuts.*, vol. 19, no. 2, pp. 828–854, 2nd Quart., 2017.

[9] I. B. Prasad, Yogita, S. S. Yadav, and V. Pal, "HLBC: A hierarchical layer-balanced clustering scheme for energy efficient wireless sensor networks," *IEEE Sensors J.*, vol. 21, no. 22, pp. 26149–26160, Nov. 2021.

[10] S. Kassan, J. Gaber, and P. Lorenz, "Game theory based distributed clustering approach to maximize wireless sensors network lifetime," *J. Netw. Comput. Appl.*, vol. 123, pp. 80–88, Dec. 2018.

[11] N. Xing, Q. Zong, L. Dou, B. Tian, and Q. Wang, "A game theoretic approach for mobility prediction clustering in unmanned aerial vehicle networks," *IEEE Trans. Veh. Technol.*, vol. 68, no. 10, pp. 9963–9973, Oct. 2019.

[12] G. Koltsidas and F.-N. Pavlidou, "A game theoretical approach to clustering of ad-hoc and sensor networks," *Telecommun. Syst.*, vol. 47, nos. 1–2, pp. 81–93, Jun. 2011.

[13] X. Chen, Y. Yin, and Z. Xu, "A game-theoretic approach for efficient clustering in wireless sensor networks," in *Proc. Int. Conf. Comput. Inf. Sci.*, Jun. 2013, pp. 1663–1666, doi: 10.1109/ICCIS.2013.435.

[14] D. Xie, Q. Sun, Q. Zhou, Y. Qiu, and X. Yuan, "An efficient clustering protocol for wireless sensor networks based on localized game theoretical approach," *Int. J. Distrib. Sensor Netw.*, vol. 9, no. 8, pp. 264–273, 2013.

[15] L. Yang, Y.-Z. Lu, Y.-C. Zhong, X.-G. Wu, and S.-J. Xing, "A hybrid, game theory based, and distributed clustering protocol for wireless sensor networks," *Wireless Netw.*, vol. 22, no. 3, pp. 1007–1021, Apr. 2016.

[16] Q. Liu and M. Liu, "Energy-efficient clustering algorithm based on game theory for wireless sensor networks," *Int. J. Distrib. Sensor Netw.*, vol. 13, no. 11, Nov. 2017, Art. no. 155014771774370.

[17] W. B. Heinzelman, A. P. Chandrakasan, and H. Balakrishnan, "An application-specific protocol architecture for wireless microsensor networks," *IEEE Trans. Wireless Commun.*, vol. 1, no. 4, pp. 660–670, Oct. 2002.

[18] E. E. Tsiropoulou, S. T. Paruchuri, and J. S. Baras, "Interest, energy and physical-aware coalition formation and resource allocation in smart IoT applications," in *Proc. 51st Annu. Conf. Inf. Sci. Syst. (CISS)*, 2017, pp. 1–6.

[19] K. Kojima, D. Fudenberg, and J. Tirole, "Game Theory," *J. Econ. Bus. Admin.*, vol. 166, 1992.

[20] L. Cao, Y. Yue, Y. Cai, and Y. Zhang, "A novel coverage optimization strategy for heterogeneous wireless sensor networks based on connectivity and reliability," *IEEE Access*, vol. 9, pp. 18424–18442, 2021.

[21] M. Zou, Z. Ping, S. Zheng, X. Qin, and H. Tongzhi, "A novel energy efficient converage control in WSNs based on ant colony optimization," in *Proc. Int. Symp. Comput., Commun., Control Autom. (3CA)*, May 2010, pp. 523–527.

[22] J. Kennedy and R. Eberhart, "Particle swarm optimization," in *Proc. Int. Conf. Neural Netw. (ICNN)*, vol. 4. Perth, WA, Australia, Nov./Dec. 1995, pp. 1942–1948.

[23] B. M. Sahoo, H. M. Pandey, and T. Amgoth, "GAPSO-H: A hybrid approach towards optimizing the cluster based routing in wireless sensor network," *Swarm Evol. Comput.*, vol. 60, Feb. 2021, Art. no. 100772.

[24] M. Y. Arafat and S. Moh, "Localization and clustering based on swarm intelligence in UAV networks for emergency communications," *IEEE Internet Things J.*, vol. 6, no. 5, pp. 8958–8976, Oct. 2019.

[25] S. Arjunan and P. Sujatha, "Lifetime maximization of wireless sensor network using fuzzy based unequal clustering and ACO based routing hybrid protocol," *Appl. Intell.*, vol. 48, no. 8, pp. 2229–2246, 2017.

[26] T. M. Behera, S. K. Mohapatra, U. C. Samal, M. S. Khan, M. Daneshmand, and A. H. Gandomi, "Residual energy-based cluster-head selection in WSNs for IoT application," *IEEE Internet Things J.*, vol. 6, no. 3, pp. 5132–5139, Jun. 2019.

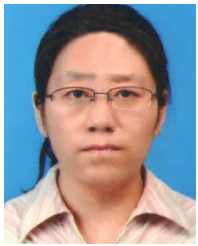

**Yin-Di Yao** was born in Baoji, China, in April 1978. She received the M.S. degree in communications and information systems. She is a Senior Engineer with the Xi'an University of Posts and Telecommunications, Xi'an, China. Her current research interests include the technology and application of the Internet of Things.

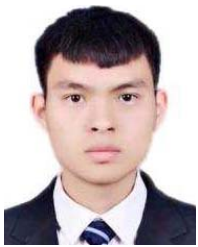

**Xiong Li** (Student Member, IEEE) is pursuing the master's degree with the Xi'an University of Posts and Telecommunications, Xi'an, China. He won the "Featured Article" Award in IEEE SENSORS JOURNAL. His research interests include the technology and application of the Internet of Things, semantic communication, and industrial internet.

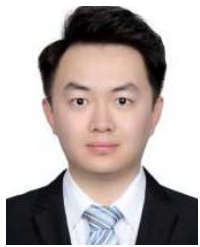

**Yan-Peng Cui** (Student Member, IEEE) received the M.S. degree from the Xi'an University of Posts and Telecommunications, Xi'an, China, in 2020. He is currently pursuing the Ph.D. degree with the School of Information and Communication Engineering, Beijing University of Posts and Telecommunications. His research interests include the technology and application of the Internet of Things and UAV networks.

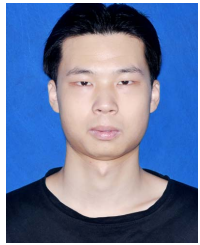

**Lang Deng** is pursuing the bachelor's degree with the Xi'an University of Posts and Telecommunications, Xi'an, China. His research interests include the technology and application of the Internet of Things.

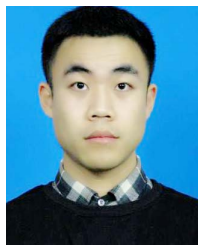

**Chen Wang** is pursuing the master's degree with the Xi'an University of Posts and Telecommunications, Xi'an, China. His research interests include the technology and application of the Internet of Things.